\documentclass[conference]{IEEEtran}
\IEEEoverridecommandlockouts
\usepackage{cite}
\usepackage{amsmath,amssymb,amsfonts}
\usepackage{algorithmic}
\usepackage{graphicx}
\usepackage{textcomp}
\usepackage{xcolor}
\usepackage{tabularx}
\usepackage{datatool}
\usepackage{float}
\usepackage{booktabs}

\def\BibTeX{{\rm B\kern-.05em{\sc i\kern-.025em b}\kern-.08em
    T\kern-.1667em\lower.7ex\hbox{E}\kern-.125emX}}

\begin{document}


\thispagestyle{empty}

\begin{huge}
IEEE Copyright Notice
\end{huge}

\vspace{5mm} 

\vspace{5mm} 

\begin{large}
© 2023 IEEE.  Personal use of this material is permitted.  Permission from IEEE must be obtained for all other uses, in any current or future media, including reprinting/republishing this material for advertising or promotional purposes, creating new collective works, for resale or redistribution to servers or lists, or reuse of any copyrighted component of this work in other works.
\end{large}

\vspace{5mm} 

\begin{large}
\textbf{Presented at:} The 2023 World Congress in Computer Science, Computer Engineering, and Applied Computing (CSCE'23) July 24-27, 2023, Luxor (MGM), Las Vegas, USA
https://american-cse.org/csce2023/
\end{large}

\vspace{5mm} 


\newcolumntype{L}[1]{>{\raggedright\arraybackslash}p{#1}}
\newcolumntype{C}[1]{>{\centering\arraybackslash}p{#1}}
\newcolumntype{R}[1]{>{\raggedleft\arraybackslash}p{#1}}

\clearpage
\pagenumbering{arabic} 

\title{Not as Simple as It Looked: Are We Concluding for Biased Arrest Practices?}

\makeatletter
\newcommand{\linebreakand}{
  \end{@IEEEauthorhalign}
  \hfill\mbox{}\par
  \mbox{}\hfill\begin{@IEEEauthorhalign}
}

\makeatother
\author{
  \IEEEauthorblockN{Murat Ozer}
  \IEEEauthorblockA{\textit{School of Information Technology} \\
    \textit{University of Cincinnati}\\
    Cincinnati, Ohio, USA \\
    m.ozer@uc.edu}
  \and
  \IEEEauthorblockN{Halil Akbas}
  \IEEEauthorblockA{\textit{Criminal Justice Department} \\
    \textit{Troy University}\\
    Troy, AL, USA \\
    hakbas@troy.edu}
  \and
  \IEEEauthorblockN{Ismail Onat}
  \IEEEauthorblockA{\textit{Criminal Justice and Criminology Department} \\
    \textit{University of Scranton}\\
    Scranton, PA, USA \\
    ismail.onat@scranton.edu}
  \and
  \IEEEauthorblockN{Mehmet Bastug}
  \IEEEauthorblockA{\textit{CJ and Criminology Department} \\
    \textit{University of Scranton}\\
    Scranton, PA, USA \\
    mehmet.bastug@scranton.edu}
  \and
  \IEEEauthorblockN{Arif Akgul}
  \IEEEauthorblockA{\textit{School of Criminology} \\
    \textit{Indiana State University}\\
    Cincinnati, Ohio, USA \\
    Arif.Akgul@indstate.edu}
  \and
  \IEEEauthorblockN{Nelly Elsayed}
  \IEEEauthorblockA{\textit{School of Information Technology} \\
    \textit{University of Cincinnati}\\
    Cincinnati, Ohio, USA \\
    elsayeny@ucmail.uc.edu}
  \and
  \IEEEauthorblockN{Zag ElSayed}
  \IEEEauthorblockA{\textit{School of Information Technology} \\
    \textit{University of Cincinnati}\\
    Cincinnati, Ohio, USA \\
    elsayezs@ucmail.uc.edu}
      \and
  \IEEEauthorblockN{Multu Koseli}
  \IEEEauthorblockA{\textit{Political Science} \\
    \textit{Chicago State University}\\
    Chicago, IL, USA \\
    mkoseli@csu.edu}
          \and
  \IEEEauthorblockN{Niyazi Ekici}
  \IEEEauthorblockA{\textit{Law Enforcement and Justice} \\
    \textit{Western Illinois University}\\
    Macomb, IL, USA \\
    N-Ekici@wiu.edu }
}

\maketitle

\thispagestyle{plain}
\pagestyle{plain}

\begin{abstract}

This study examines racial disparities in violent arrest outcomes, challenging conventional methods through a nuanced analysis of Cincinnati Police Department data. Acknowledging the intricate nature of racial disparity, the study categorizes explanations into types of place, types of person, and a combination of both, emphasizing the impact of neighborhood characteristics on crime distribution and police deployment. By introducing alternative scenarios, such as spuriousness, directed policing, and the geo-concentration of racial groups, the study underscores the complexity of racial disparity calculations. Employing a case study approach, the analysis of violent arrest outcomes reveals approximately 40 percent of the observed variation attributed to neighborhood-level characteristics, with concentrated disadvantage neutralizing the influence of race on arrest rates. 

Contrary to expectations, the study challenges the notion of unintentional racism, suggesting that neighborhood factors play a more significant role than the racial composition in explaining arrests. Policymakers are urged to focus on comprehensive community development initiatives addressing socioeconomic inequalities and support the development of robust racial disparity indices. The study calls for nuanced explorations of unintentional racism and future research addressing potential limitations, aiming to enhance understanding of the complexities surrounding racial disparities in arrests.
\end{abstract}
 
\section{Introduction}

Racial disparity in police activities, encompassing stops, citations, and arrests, is a contentious and widely discussed issue, especially in the aftermath of tragic events like Ferguson and George Floyd. Public attention, heightened by such incidents, often centers on racialized police actions. News statistics frequently highlight the disproportionate targeting of minority groups by the police, employing simplistic calculations that divide stops, searches, or arrests by the corresponding population of each racial group and then compare the rates.

Empirical studies face challenges similar to the above approach, partly due to low-quality data lacking contextual details. Despite the complexity of denominators, scholars often simplify the calculation of racial disparity into a single denominator, a method criticized for its potential harm to law enforcement agencies \cite{neil2019methodological}. This conventional approach suggests differential treatment of racial groups by the police.
Existing knowledge on racial disparity primarily stems from past studies focusing on a single dimension within limited settings, while studies exploring alternative scenarios are rare. In this context, our study revisits alternative scenarios of racial disparity, including confounding factors, directing policing, geo-concentration of racial groups, gang territories, prolific offenders, and the ecological fallacy. This exploration challenges current understandings of racial disparity.

The study unfolds as follows: first, we acknowledge the intricate nature of racial disparity. Second, we revisit racial disparity outcomes, highlighting the pitfalls of using simplified denominators for benchmarking. Third, utilizing arrest data from the Cincinnati Police Department (CPD), we introduce a nuanced racial disparity formula, offering alternative scenarios to demonstrate the complexity of arrest outcomes. Subsequently, we employ a hierarchical linear regression model, incorporating multi-dimensional settings to assess how apparent racial disparity results can be mitigated. The conclusion section discusses policy implications derived from our study.

\section{Complex Nature of Racial Disparity}

The challenge of selecting a baseline measure for calculating racial disparities in criminal justice outcomes is complex and context-dependent, with variations across different types of criminal justice outcomes such as violent crime arrests, drug-related arrests, traffic stops, and stop-and-frisk approaches. The literature indicates that racial disparity outcomes result from intricate and reciprocal factors, making it difficult to determine a suitable baseline measure. While this study does not delve into this issue extensively, a brief summary of the literature reveals three main categories of explanations: (a) types of place, (b) types of person, and (c) a combination of both.

Types of Place Explanations: Shaw and McKay's work (1942) emphasizes that crime is influenced by factors like poverty, residential mobility, and ethnic heterogeneity concentrated in inner-city zones. The non-random distribution of crime, particularly in hot spots, aligns with structural inequalities prevalent in areas where most Black individuals reside. This situation, described by Shaw and McKay (1949), persists over the years and influences police deployment patterns, leading to a higher likelihood of encountering Black individuals and subsequent arrests \cite{engel2012race}\cite{warren2006driving}.

Types of Person Explanations: Criminological theories highlight correlations between family structure, subcultures of violence, and deprivation/poverty with race. Merton's strain theory argues that individuals with limited legitimate goals may resort to criminal or deviant acts. Cloward and Ohlin (1960) suggest that communities with limited opportunities for conventional behaviors are more prone to deviant adaptations, influenced by peer networks promoting criminogenic behavior \cite{cloward2013delinquency}.

Combination of Place and Person Explanations: Studies propose a complex model synthesizing both types of place and types of person explanations. Concentrating police efforts in crime-intense locations, often Black-populated inner-city zones, can lead to negative citizen-police interactions, fostering resentment and antagonistic behaviors. This, in turn, increases the likelihood of arrests \cite{engel2000further}\cite{terrill2007nonarrest}\cite{worden1996demeanor}\cite{brown2011racial}.

\section{Disparate Outcomes of Racial Disparity}
Studies examining racial disparities in police outcomes, such as traffic stops, citations, searches, and warnings, employ various baseline measures or benchmarking variables. Initial studies using the general population as a baseline measure found substantial racial disparities in traffic stops \cite{lamberth1996report}\cite{verniero1999interim}. More recent studies utilize better baseline variables, such as roadway observations, to determine drivers' volumes of specific racial groups, resulting in more reliable racial disparities \cite{meehan2002race}\cite{engel2006cleveland}\cite{lamberth2006data}\cite{farrel2006rhode}.

The theme of racial disparities in police arrests, gaining prominence due to media reports and national statistics, attracts attention. While some argue that arrest decisions align with racial distributions in offending districts, others suggest that employing the frequency of crime involvement as a baseline measure yields different findings. The complexity of identifying a suitable baseline measure is evident in studies comparing drug arrest rates between Blacks and Whites based on the frequency of drug users and deliverers. Baseline measure choices, such as needle exchange survey data or citizen-initiated calls for service (CFS) data, significantly impact the calculated racial disparities in criminal justice outcomes \cite{beckett2006race}\cite{engel2012race}.

\section{Alternative Scenarios}
The study by Pierson et al. (2020), analyzing a vast dataset of traffic stops, revealed a higher likelihood of African Americans being stopped compared to their white counterparts \cite{pierson2020large}. However, relying solely on a simplified racial disparity calculation raises concerns about internal and external validity. Alternative explanations should be considered:
Spuriousness: Simplifying variables into a single denominator for racial disparity calculations risks omitting relevant factors, potentially leading to spurious relationships. This method overlooks essential contextual and behavioral variables, compromising research integrity \cite{neil2019methodological}.

Directed Policing: Criminological theories emphasize crime hotspots, where a small percentage of locations contribute to the majority of crimes. Law enforcement agencies strategically focus on these areas to reduce crime effectively using GIS technologies. A benchmark denominator fails to account for directed policing strategies, impacting the interpretation of racial disparity outcomes \cite{briggs2017impact}\cite{wheeler2020allocating}.
\textit{Geo-Concentration of Racial Groups:} Population-based benchmark variables ignore the spatial concentration of racial groups within a given geography. For instance, if one block group has an equal number of Blacks and Whites but the Black population is concentrated in specific blocks, the police are more likely to encounter and interact with Blacks, skewing racial disparity perceptions.

Gang Turfs and Prolific Offenders: Studies suggest that a small fraction of the population, particularly gang members, is responsible for a significant portion of crimes. Focused deterrence approaches targeting gang-controlled territories may result in racially disproportionate outcomes. Assuming discriminatory policing based on arrest numbers without considering the prevalence of prolific offenders can be misleading \cite{engel2013reducing}.
\textit{Ecological Fallacy:} Structural inequalities, as explained by Shaw and McKay (1942), contribute to crime-dense locations irrespective of race. Aggregating outcomes like stops, searches, and arrests to the city level may mask individual block group differences, potentially leading to misconceptions of racially biased outcomes \cite{shaw1942juvenile}.
In summary, while Pierson et al.'s study sheds light on racial disparities in traffic stops, acknowledging and addressing the limitations associated with simplified racial disparity calculations is crucial for a nuanced understanding of the findings.

\section{ Case Study: Exploring Racial Disparity In Violent Arrest Outcomes}

This section of the study delves into the analysis of racial disparity in violent arrest outcomes using Cincinnati Police Department data, focusing on homicides, rape, robbery, and felonious assault. The choice of violent crimes is strategic, as police discretion is minimal in these cases \cite{goldstein1977policing}\cite{engel2002theory}, ensuring a controlled study of arrest outcomes. Warrant arrests, resulting from police encounters with criminals, constitute the majority of cases.

Initially, we employed population as the benchmark denominator to assess the ratio of Black arrestees to White arrestees. To enhance this analysis, we followed Brantingham et al.'s (2019) recommendation by incorporating the number of crimes committed by each racial group. This step aims to consider differential crime commission rates among racial groups. Subsequently, as suggested by Neil and Winship (2019), we applied sophisticated data analysis techniques to scrutinize the validity of inferences derived from the initial simplified analysis (benchmark). This comprehensive approach ensures a nuanced understanding of racial disparity in violent arrest outcomes.

\vspace{1em} 
\textbf{Data} 

In our study, we utilized six diverse data sources to conduct a comprehensive analysis. The first dataset comprises reported crime records, documenting incident details, location, and the Uniform Crime Report (UCR) code indicating the seriousness of offenses. This dataset served as the basis for obtaining counts of violent crimes occurring between January 1, 2011, and December 31, 2015.

The second data source involves incident-level suspect data covering the period from January 1, 2011, to December 31, 2015. This dataset includes information on suspects linked to reported crimes, featuring demographics such as sex, race, and age. Our purpose in using this dataset was to calculate the varying rates of crime commission among different racial groups, specifically Blacks and Whites.

The third dataset pertains to arrest data, encompassing records of arrestees resulting from various police activities, including traffic stops, calls for service, and search warrants. This dataset includes demographic information on arrestees and the seriousness of offense codes (Uniform Crime Report). We leveraged this dataset to derive counts of violent arrests for Whites and Blacks during the period from January 1, 2011, to December 31, 2015.

Our fourth data source was derived from the 2010 Census data, wherein we obtained block group information for the State of Ohio using the Esri Tiger webpage. Through ArcMap 10.1 software, we narrowed down the data to the City of Cincinnati, focusing on pertinent variables (refer to Table 1 for descriptive statistics).

The fifth dataset was sourced from the "2010 Cincinnati Statistical Neighborhood Approximations," utilizing Census Tract data to extract neighborhood-level covariates like population and income levels. Cincinnati's neighborhoods (N=50) were well-defined based on historical roots, and the statistical neighborhood approximation logarithm was employed to align Census Tract data with the city's historic neighborhoods. This neighborhood-level data was instrumental in exploring the influence of block group racial characteristics in structurally distinct neighborhoods.

Our final data points were derived from venues with liquor licenses, sourced from the Cincinnati Police Department data files. Utilizing ArcGIS 10.1, we aggregated liquor store addresses to form block groups, obtaining the count of liquor stores in each block group. This inclusion was guided by prior research indicating a significant association between the concentration of liquor stores and crime hot spots (see Block and Block, 1995; Bernasco and Block, 2011) \cite{bullock1955urban}\cite{bernasco2005residential}.

\vspace{1em} 
\textbf{Unit of Analysis and Data Preparation}

In our data compilation, we adopted two distinct units of analysis: the block group level and the neighborhood level. A longstanding debate surrounds the ideal size of the analysis unit, with varying opinions on whether larger units, such as city-level comparisons, might overshadow the nuances observed at smaller units, potentially masking or revealing racial disparities (as discussed in the concept of ecological fallacy).

To ensure a detailed examination and to address anonymity concerns, we sought racial compositions at the smallest available unit: the block group level in Census data. Leveraging Ohio State's block group data from the Esri Tiger webpage, we refined this data within ArcMap 10.1 for the City of Cincinnati. Notably, we identified 291 block groups within Cincinnati and selected specific block group covariates for our study (refer to Table 1).

The block group level served as our primary unit of analysis due to its granularity and detailed demographic reporting. Simultaneously, we incorporated neighborhood-level variables as an upper-level unit of analysis. This approach aimed to explore the variations in the relationship between block group covariates and our outcome measure concerning diverse contexts of neighborhood-level characteristics.

\vspace{1em} 
\textbf{Measures of Variables}

In line with previous research indicating that police exercise greater discretion for less severe offenses, potentially revealing any disproportionate racial distribution more prominently in minor offenses \cite{goldstein1977policing}\cite{engel2002theory}, we chose to focus on the number of violent crime arrests as our outcome variable. This decision aimed to mitigate the influence of confounding variables on police discretion, aligning with the insights from retrospective studies.

As illustrated in Table 1, the distribution of the outcome variable exhibits substantial skewness, a condition incompatible with linear regression analysis. To address this, we applied a logarithmic transformation to the outcome variable, effectively addressing the skewness issue and ensuring a normal distribution of the logged dependent variable.

\begin{table*}[h]
\centering
\caption{Descriptive Statistics}
\label{tab:descriptive_statistics}
\begin{tabular}{lcccc}
\hline
\multicolumn{5}{c}{Dependent Variables} \\ \hline
 & N & Min-Max & Mean & s.d. \\
Number of Violent Arrest & 291 & 0 - 163 & 14.28 & 16.91 \\
Number of Violent Arrest (logged) & 291 & 0 - 5.10 & 2.22 & 1.09 \\ \hline
\multicolumn{5}{c}{Independent Variables} \\ \hline
\multicolumn{5}{c}{Block Group Level Descriptive Statistics} \\
 & N & Mean & s.d. & Min - Max \\
Population & 291 & 1082.1 & 513.52 & 191 - 3736 \\
Black Population & 291 & 455.2 & 452.94 & 0 - 3415 \\
Number of Violent Crime Black Suspects & 291 & 47.99 & 51.56 & 0 - 341 \\
Number of Violent Crime White Suspects & 291 & 4.62 & 6.2 & 0 - 50 \\
Percent Female-Headed Households & 291 & 14.83 & 11.01 & 0 - 66.56 \\
Percent Households Under Poverty & 291 & 39.42 & 40.15 & 0 - 65.07 \\ \hline
\multicolumn{5}{c}{Neighborhood Level Descriptive Statistics} \\
 & N & Mean & s.d. & Min - Max \\
Number of Liquor Stores & 50 & 15.74 & 18.3 & 0 - 112 \\
Percent Vacant Houses & 50 & 21.8 & 9.95 & 6.35 - 44.74 \\
Concentrated Disadvantage & 50 & 0 & 1 & -5.79 \\ \hline
\multicolumn{5}{c}{Factor Loadings of Neighborhood Level Concentrated Disadvantage} \\
Percent Female-Headed Households & & & & 0.885 \\
Percent Violent Crime & & & & 0.811 \\
Percent Households Living Under Poverty & & & & 0.797 \\
Percent Rental Properties & & & & 0.78 \\ \hline
\multicolumn{5}{c}{KMO and Barlett's test=.696; Determinant=.174} \\ \hline
\end{tabular}
\end{table*}

\vspace{1em} 
\textbf{Benchmark Comparison to Calculate Racial Disparity}

In our study, we incorporated the variable of black population to investigate the relationship between race (Blacks) and the examined outcome variable. Our hypothesis posited that Blacks experience disproportionate arrests compared to Whites. To standardize racial compositions across block groups, we utilized the population of each block group, specifically focusing on the Black population.

Table 2 presents the analysis of the differential violent crime commission between Blacks and Whites. Initially, we computed racial disparity, as outlined in Equation 1, utilizing population as the benchmark variable, a conventional approach employed by numerous studies in the past.
\begin{equation}
\text{Traditional Formula} = \frac{\left(\frac{\text{\# of Black Arrestees}}{\text{Black Population}}\right)}{\left(\frac{\text{\# of White Arrestees}}{\text{White Population}}\right)}
\end{equation}

The outcome of the traditional equation is presented in the "Disparity index and Population Benchmark" column of Table 2. According to the findings, Blacks are, on average, six times more likely to be arrested compared to their White counterparts. Moreover, our study incorporates the factor of differential crime involvement for these racial groups into the equation, as demonstrated in Equation 2.

\begin{equation}
\frac{\frac{\text{\# of Black Arrestees}}{\text{Black Population}}}{\frac{\text{\# of White Arrestees}}{\text{White Population}}} \Bigg/ \frac{\frac{\text{V.Crimes Commission of Blacks}}{\text{Black Pop.}}}{\frac{\text{V.Crimes Commission of Whites}}{\text{White Pop.}}}
\end{equation}

This approach aligns with the methodology employed in the recent study by Bratingham et al. (2018). Consequently, upon incorporating the differential crime commission into the calculation, the racial disparity gap witnessed a significant reduction, as illustrated in the last column of Table 2. The desired outcome for racial disparity is one (1), indicating no racial disparity. Calculating Blacks' rate to Whites, any value exceeding one implies that Blacks were more likely to be arrested, considering their representation in the population and the difference in crime involvement. Table 2 indicates that Blacks are less likely to be arrested when accounting for their population and the rates of differential crime involvement.

\begin{table*}[htbp]
  \centering
  \caption{Disparity Indexes for the Outcome Variables}
  \begin{tabular}{cccccccccc}
    \toprule
    Year & \shortstack{Violent Crimes\\Committed by\\Blacks} & \shortstack{Violent Crimes\\Committed by\\Whites} & \# of Black Arrestees & \# of White Arrestees & \shortstack{Black\\Population} & \shortstack{White\\Population} & \shortstack{Disparity\\Index \&\\Population\\Benchmark} & \shortstack{Disparity\\Index,\\Population \&\\Differential\\Involvement} \\
    \midrule
    2015 & 2492 & 288 & 616 & 113 & 133,869 & 147,315 & 6 & 0.63 \\
    2014 & 2486 & 288 & 671 & 107 & 133,578 & 146,995 & 6.9 & 0.73 \\
    2013 & 2675 & 231 & 688 & 115 & 133,288 & 146,676 & 6.58 & 0.52 \\
    2012 & 3272 & 315 & 776 & 144 & 132,854 & 146,199 & 5.93 & 0.52 \\
    2011 & 3144 & 248 & 814 & 187 & 132,613 & 145,933 & 4.79 & 0.34 \\
    \bottomrule
  \end{tabular}
\end{table*}

\vspace{1em} 
\textbf{Replicating Racial Disparity with the Regression Analysis}

To explore the relationship between arrests and crimes, proxy variables reflecting neighborhood characteristics were employed, including the percent of female-headed households and the percent of households living under poverty at the block group level. These variables align with social disorganization theory and previous findings linking them to higher crime rates \cite{dornbusch1985single}\cite{kawachi1999crime}\cite{mclanahan1989mother}\cite{harding2010living}\cite{krivo1996extremely}\cite{sampson1997racial}. 

Additionally, the number of Black suspects was included as an independent variable at the block group level to address the notion that disproportionate arrest rates may be influenced by differential crime commissions among racial groups \cite{dalessio2003race}\cite{hindelang1978race}.

At the neighborhood level, three variables were considered: the number of liquor stores, the percent of vacant houses, and a measure of concentrated disadvantage derived from neighborhood-level indicators. Research has established connections between the concentration of liquor stores and crime, as well as the impact of vacant houses on neighborhood deterioration and criminal activities \cite{conrow2015spatio}\cite{gorman2015spatial}\cite{lipton2013geography}\cite{white2012alcohol}\cite{zhu2004alcohol}\cite{hannon2006neighborhood}\cite{spelman1993abandoned}. The concentrated disadvantage measure amalgamated indicators such as the percent of female-headed households, percent rental properties, percent violent crimes, and percent households living under the poverty line.

Prior to multivariate analysis, a basic comparison was made by comparing Blacks' and Whites' arrest counts for violent crimes, utilizing their corresponding population estimates as a benchmark (see Equation 1). However, this basic formula has limitations, such as overlooking local-level racial disparity outcomes. To address this, multivariate analysis was employed, utilizing block group covariates at the micro-level and neighborhood-level indicators at the macro level. The study utilized a Hierarchical Linear Model to maintain independent error terms for block groups \cite{raudenbush2003hierarchical}. Group-mean centering was applied to examine the effects of block group variables and level-2 variables separately, allowing the assessment of a block group's relative position within the group \cite{hox2002multilevel}. The intercept in the analysis represents the mean level of violent crime arrests when level-1 predictors are set equal to their corresponding neighborhood unit means

\vspace{1em} 
\textbf{Multivariate Analysis}

The racial disparity variable calculation, as previously discussed, is intricate, often requiring consideration of various factors. Multivariate analysis serves as a crucial tool for comprehending the social and economic determinants behind the racial disparity gap, providing valuable insights for jurisdictions. In this study, we employed the Hierarchical Linear Model (HLM) to confirm the racial disparity calculation and understand the underlying factors influencing it.

The "unconditional model" of the HLM analysis revealed a significant variation in average violent crime arrests across the 50 Cincinnati neighborhoods (t ratio= 19.09, reliability estimate of violent crime coefficient= 0.705). The interclass correlation coefficient (ICC) indicated a 41.48\% variation at the neighborhood level. Subsequently, three neighborhood-level variables—number of liquor stores, percent vacant houses, and concentrated disadvantage—were identified as the most effective level-2 predictors to explain the substantial variance observed in the unconditional model.

Having identified the appropriate level-2 variables, both level-1 and level-2 variables were introduced into the HLM equation. Model 1 of Table 3 exclusively incorporates level-1 variables. With the exception of predictors like population and percent female-headed household, all level-1 predictors significantly correlated with the outcome variable. Specifically, block groups with evident poverty were more prone to experiencing violent crime arrests. Notably, even with the control of violent crime suspects at the block group level, the race (Blacks) variable retained a significant association with violent arrest outcomes. This finding implies that Blacks were disproportionately arrested for violent crimes, regardless of their corresponding population and violent crime commission rate. It's worth noting that the "Black population" slope was specified as random to assess whether the race variable exhibited consistent effects across neighborhoods with varying characteristics, such as low concentrated disadvantage vs. high concentrated disadvantage.

\begin{table*}[htbp]
  \centering
  \caption{HLM OLS Regression Model for Violent Crime Arrests}
  \begin{tabular}{lcccccccc}
    \toprule
    & \multicolumn{3}{c}{Model 1} & \multicolumn{3}{c}{Model 2} \\
    
    Fixed Effects & $b$ & $se$ & $p$-value & $b$ & $se$ & $p$-value \\
    \hline
    Mean Violent Crime Arrest (base) & 2.31 & 0.122 & 0.000 & 2.02 & 0.129 & 0.000 \\
    Neighborhood Level Percent Vacant & - & - & - & 0.015 & 0.002 & 0.000 \\
    Neighborhood-level Liquor Stores & - & - & - & 0.017 & 0.003 & 0.000 \\
    Neighborhood-level Concentrated Disadvantage & - & - & - & 0.959 & 0.101 & 0.000 \\
    Spatial Autocorrelation Lag & -0.012 & 0.097 & 0.904 & -0.005 & 0.098 & 0.960 \\
    \hline
    Alternative Neighborhood Level $R^2$ & - & 0.355 \\
    \hline
    Percent Female Headed Householders & -0.005 & 0.006 & 0.372 & -0.0005 & 0.006 & 0.407 \\
    Percent Householders Living under Poverty & 0.0036 & 0.002 & 0.026 & 0.0035 & 0.002 & 0.033 \\
    Number of Black Suspects of Violent Crimes & 0.0116 & 0.002 & 0.000 & 0.0125 & 0.002 & 0.000 \\
    Black Population (base) & 0.0005 & 0.0002 & 0.024 & -0.00004 & 0.0006 & 0.950 \\
    Neighborhood Level Percent Vacant & - & - & - & 0.00001 & 0.00002 & 0.609 \\
    Neighborhood-level Liquor Stores & - & - & - & 0.00001 & 0.00002 & 0.457 \\
    Neighborhood-level Concentrated Disadvantage & - & - & - & -0.0005 & 0.0002 & 0.006 \\
    Overall Population & -0.0002 & 0.0001 & 0.209 & -0.0002 & 0.0001 & 0.284 \\
    \hline
    Alternative Street Level $R^2$ & 0.368 & 0.392 \\
    \hline
    Random Effects & \multicolumn{2}{c}{Variance Component} & \multicolumn{2}{c}{Variance Component} \\
    Level-2 & 0.61287 & 0.07491 \\
    Level-1 & 0.46055 & 0.44372 \\
    \hline
    Analysis is based on 291 block groups and 50 neighborhoods \\
    \bottomrule
  \end{tabular}
\end{table*}

In light of this context, Model 2 in Table 3 incorporates neighborhood covariates. All the main effects of neighborhood-level variables significantly contribute to explaining the mean level of violent crime arrests. Specifically, violent crime arrests are more likely to occur in neighborhoods characterized by a high number of liquor stores, a high percentage of vacant houses, and prevalent concentrated disadvantage. With the exception of race (Black population), all level-1 predictors remain consistent in Model 2. Additionally, a spatial lag variable was introduced to assess whether spatial autocorrelation influences arrest outcomes, aiming to control for the potential impact of nearby areas on disparate arrests in a given block group. Results across different tables indicate that spatial autocorrelation's influence is negligible.

In Model 2 of Table 3, only concentrated disadvantage exhibits a significant and negative interaction with Blacks' random slope. The negative sign of neighborhood covariates implies that the concentrated disadvantage variable diminishes or mitigates the effect of the block group race (Blacks) variable on violent crime arrests. More explicitly, concentrated disadvantage entirely nullifies or neutralizes the impact of race on violent crime arrests. In other words, being Black does not predict violent crime arrests when economic and social deprivations within communities are considered in the equation.	

\section{Discussion and Conclusion}

The findings of this study underscore the prevalent issue of racial disparity in arrest outcomes, confirming that, akin to existing research, Blacks face a higher likelihood of arrest compared to their White counterparts in the City of Cincinnati. The analysis, focused on violent crime arrests, delved into the mediating role of covariates derived from hot spot policing, strain, and social disorganization theories. Approximately 40 percent of the observed variation in violent crime arrests was attributed to neighborhood-level characteristics. Notably, the race variable exhibited significant variation across Cincinnati neighborhoods, suggesting nuanced arrest patterns for Blacks in different areas. The introduction of neighborhood-level attributes demonstrated their impact on both the intercept of violent crime arrests and the race variable's slope.

The cross-level interactions revealed a crucial policy implication: the socioeconomic deprivations of neighborhoods, specifically concentrated disadvantage, effectively neutralized the influence of race on violent crime arrest rates. This implies that addressing neighborhood-level disadvantages, rather than solely focusing on racial disparities, could be a key strategy in reducing disproportionate arrest rates. Policymakers should consider comprehensive community development initiatives targeting socioeconomic inequalities and neighborhood-level challenges to tackle the root causes of disparate arrest outcomes.

These findings align with social disorganization theory, affirming the association between Black populations, concentrated disadvantage, and higher crime rates \cite{shaw1942juvenile}. The study emphasizes the robust predictive power of neighborhood-level concentrated disadvantage, eclipsing the influence of race on violent arrests.

Furthermore, the study incorporates three neighborhood-level variables (concentrated disadvantage, liquor stores, and vacant houses), recognized as prime predictors of crime concentration or hot spots. The significant associations of these characteristics with arrest outcomes underscore the importance of place-based factors in understanding disproportionate arrest rates. This supports the growing body of literature emphasizing the role of place characteristics in shaping arrest patterns.

The study suggests that strategic deployment based on place characteristics, as identified through hot spot analysis, could be a valuable approach for law enforcement agencies. Policymakers should encourage police departments to leverage analytical capabilities to target hot spots effectively, acknowledging the significance of neighborhood-level variables in explaining arrest disparities. However, caution is warranted in extrapolating these findings to assert intentional racism, urging future studies to scrutinize deployment patterns and their potential unintended consequences.

Contrary to expectations, the study challenges the notion of "unintentional racism" resulting from non-random deployment strategies. The findings indicate that concentrated disadvantage and other neighborhood factors, rather than the percent Black population, primarily explain differential arrests in Cincinnati neighborhoods. This raises questions about the conventional wisdom regarding unintentional racism and highlights the need for nuanced examinations of deployment patterns.

The study prompts a nuanced exploration of the relationship between deployment patterns and unintentional racism. Policymakers should encourage further research specifically focused on hot spots to assess the extent to which agencies' deployment patterns contribute to unintended racial disparities in arrests. Understanding these dynamics can inform policy interventions aimed at minimizing unintended consequences in law enforcement strategies.

The study's most crucial contradictory finding emphasizes that calculating racial disparities in arrests is not a straightforward task, requiring a more nuanced approach beyond comparing arrest counts to population demographics. The inclusion of the differential crime involvement of racial groups is essential for accurate assessments of racial disparities in arrest outcomes.
Policymakers should advocate for the adoption of comprehensive measures that consider not only arrest counts but also the context of crime involvement by different racial groups. A thorough, detailed analysis is necessary to unravel the complex interplay of social and economic factors contributing to disproportionate arrests in a given jurisdiction. Policymakers should support research endeavors aimed at developing robust racial disparity indices that provide a more accurate and nuanced understanding of the issue.

While this study contributes valuable insights, it acknowledges certain limitations. The reliance on reported crime involvement might underestimate actual crime rates, and the study recognizes the potential impact of omitted variables related to race and crime. Policymakers should encourage future research to address these limitations, promoting the development of a more comprehensive racial disparity index that considers a broader array of factors and enhances our understanding of the complexities surrounding racial disparities in arrests.

\bibliographystyle{ieeetr}
\bibliography{references}

\end{document}